\documentclass[a4paper, amsfonts, amssymb, amsmath, reprint, showkeys, nofootinbib, twoside]{revtex4-1}
\usepackage[english]{babel}
\usepackage[utf8]{inputenc}
\usepackage[colorinlistoftodos, color=green!40, prependcaption]{todonotes}
\usepackage{mathtools}
\usepackage{amsmath}
\usepackage{amssymb}

\usepackage{amsthm}
\usepackage{mathtools}
\usepackage{physics}
\usepackage{xcolor}
\usepackage{graphicx}
\usepackage[left=23mm,right=13mm,top=35mm,columnsep=15pt]{geometry} 
\usepackage{adjustbox}
\usepackage{placeins}
\usepackage[T1]{fontenc}
\usepackage{lipsum}
\usepackage{csquotes}
\usepackage{tikz}
\usepackage{adjustbox}

\usepackage[pdftex, pdftitle={Article}, pdfauthor={Author}]{hyperref} 
\begin{document}

\title{Quantum Error Correction and the Limits of Quantum Metrology}

\author{Zhuoran Bao}
    \email[Correspondence email address: ]{zhuoran.bao@mail.utoronto.ca}
    \affiliation{Dept. of Physics, University of Toronto, Toronto, M5S 1A7, Ontario, Canada}
\author{Daniel F. V. James}
    \email[Correspondence email address: ]{
    dfv.james@utoronto.ca}
    \affiliation{Dept. of Physics, University of Toronto, Toronto, M5S 1A7, Ontario, Canada}

\date{\today}

\begin{abstract}
Quantum metrology has made amazing progress in recent decades. Researchers are always interested in finding new optimal states that improve parameter estimation. In this letter, we point out a connection between a code's error-correcting capacity and its ability to act as a sensor. We support this claim with an example relating the absorption-emission code to the sensor state for arbitrary state rotation. Such a unified theory allows one to draw inspiration from error correction to develop promising quantum sensors.
\end{abstract}
\keywords{Statistical Distance, Fisher Information, Knill-Laflamme Condition, Rotation Sensing}

\maketitle

Sensing an unknown process using specially constructed quantum states has been a topic of intense research over the past decade due to its potential applications \cite{Huver, Degen, Kaubruegger, Pezze, Reilly}. One of the most prominent theoretical findings is that by utilizing quantum correlation, one can surpass the classical shot-noise limit. Suppose we have M qubits to estimate a parameter associated with a unitary operation; classically, we use these M single qubits to perform M trials and the uncertainty in estimating the parameter is proportional to \(M^{-1/2}\). Alternatively, we can use these M qubits to create T identical initial states, each containing N quantum correlated qubits. Then, performing T trials, one would find that the uncertainty for the parameter to be proportional to \(N^{-1}T^{-1/2}\). The scaling of \(N^{-1}\) is referred to as the Heisenberg limit \cite{Giovannetti, Toth}. Compare to the shot noise limit \(M^{-1/2}=(NT)^{-1/2}\), we obtain an \(N^{-1/2}\) enhancement in scaling. In the case of sensing a polarization rotation, the second-order anti-coherent states, which often correspond to Platonic solids in their Majorana representation \cite{Goldberg2, Gunnar, Baguette}, are predicted to attain the ultimate bound called the quantum Cramer-Rao bound \cite{Cave}. Specifically, an N-photon second-order anti-coherent state, represented by a superposition of \(J=N/2\) angular momentum states, attains the quantum Cramer-Rao bound of \(4J(J+1)/3\) compared to a classical bound of \(J/2\) \cite{Goldberg}. However, identifying such states is not straightforward; one either uses the connection between the Platonic solids in Majorana representation and the second-order anti-coherent-state coefficients, or computes them by brute force using an optimization algorithm. On the other hand, great progress has been made in quantum error correction \cite{Terhal, Li, Jain, Knill}. Multiple attempts have been made to combine metrology and error correction, aiming to design sensors with error-correction capability against noise while maintaining the Heisenberg scaling \cite{Fischer, Zhou, Sahu}. Unfortunately, due to the limited capacity to perform two-qubit gates on polarization-encoded photonic qubits, there is not yet a straightforward translation of the error correction strategy into polarimetry. 

In this letter, we propose a slightly different approach to utilizing error-correction ideas. Instead of correcting errors caused by an imperfect channel, we use these ideas to simplify identification of the sensor state. We first present a brief review of quantum state distinguishability and its application to both metrology and error correction. Then, we will briefly discuss an interesting class of error-correction codes that correct angular-momentum errors in atomic state transitions: the Absorption-Emission (AE) code \cite{Jain}. These codes have been shown to be similar to the optimal state in polarization rotation sensing. We aim to identify common ground between the two strategies and draw inspiration from error-correcting codes to design new quantum sensors.

Operationally, measuring the difference between two quantum states amounts to distinguishing the probability distributions inferred from detection events. For pure states, the quantum state distinguishability is the maximal statistical distance between two multinomial distributions \cite{Wootters}. Let us first define the statistical distance of two multinomial distributions. Suppose we perform an experiment with M possible outcomes. Let us enumerate these possible outcomes with the whole numbers \(m=1,2, \dots, M\). After n trials, we will obtain an M-dimensional vector distribution \(\boldsymbol{P}\), with each component \(P_m\) being the probability of obtaining outcome m, associated with an uncertainty that depends on n. Imagine we can vary some parameter in the experiment such that a slightly different distribution \(\boldsymbol{Q}\) is obtained. The statistical distance denoted \(\Omega(\boldsymbol{P},\boldsymbol{Q})\) is a quantity that counts the number of distinguishable distributions between \(\boldsymbol{P}\) and \(\boldsymbol{Q}\) normalized by \(\sqrt{n}\) when the number of trials \(n\rightarrow\infty\). The infinitesimal form of statistical distance is the Euclidean distance of all m components, \(d\Omega=\Omega(\boldsymbol{P},\boldsymbol{P}+d\boldsymbol{P}) =\sqrt{\sum_{m=1}^M (d\Omega_m)^2}\) where \((d\Omega_m)^2 = \frac{(dP_m)^2}{4P_m}\) (See SM1). Using a change of variables to integrate, we find that the quantum distinguishability corresponds to an arc length on a unit hypersphere \( \Omega(\boldsymbol{P},\boldsymbol{Q})
= \arccos{\left(\sum_{m=1}^M\sqrt{P_mQ_m}\right)}\).
The term \(\sum_{m=1}^M\sqrt{P_mQ_m}\) is called the Bhattacharyya coefficient or the classical fidelity of a statistical distribution\cite{Bhattacharyya, Nielsen}.

The connection between quantum states and statistical distributions is established using a complete set of orthogonal projection operators. However, a single quantum state can be associated with an infinite number of distributions, since there are infinitely many set of projection operators. Thus, we need to further define the distinguishability of two states as the maximal statistical distance among all possible distributions generated by performing a projection onto a basis of the Hilbert space \cite{Wootters}:
\begin{equation}\label{distinguishability}
\begin{split}
    &\Lambda(\vert\psi\rangle,\vert\phi\rangle) = \max_{\{\boldsymbol{\hat{\Pi}\}}}\left[\Omega\left(\langle\psi\vert\boldsymbol{\hat{\Pi}}\vert\psi\rangle,\langle\phi\vert\boldsymbol{\hat{\Pi}}\vert\phi\rangle\right)\right) \\
    &= \max_{\{\boldsymbol{\hat{\Pi}}\}} \left[ \arccos{\left(\sum_{m=1}^M\sqrt{\langle\psi\vert\hat{\Pi}_m\vert\psi\rangle\langle\phi\vert\hat{\Pi}_m\vert\phi\rangle}\right)}\right].
\end{split}
\end{equation}
Here, \(\boldsymbol{\hat{\Pi}}\) is used to represent the complete set of orthogonal bases. To maximize over all basis, we notice that when decomposing the states into the projector basis \(\vert\psi\rangle=\sum_m\alpha_m\vert m\rangle\), and \(\vert\phi\rangle=\sum_m\beta_m\vert m\rangle\). The argument of the arccosine function is lower bounded by the fidelity between the states 
\(\Lambda(\vert\psi\rangle,\vert\phi\rangle)=\sum_{m=1}^M \vert\alpha_m\vert\vert\beta_m\vert\geq\left\vert\sum_{m=1}^M\alpha_m^*\beta_m\right\vert=\vert\langle\psi\vert\phi\rangle\vert\).
Now, we can define the distinguishability of two pure states as \(
    \Lambda(\vert\psi\rangle,\vert\phi\rangle) = \arccos{\left(\vert\langle\psi\vert\phi\rangle\vert\right)}\).
The set of complete projector operators that best distinguish the states is given by \(\hat{\Pi}_1=\vert\psi\rangle\langle\psi\vert\), and the other \(M-1\) orthogonal projectors satisfying \(\hat{\Pi}_m\hat{\Pi}_1=0\). Effectively, Eq. (\ref{distinguishability}) becomes \(\Lambda(\vert\psi\rangle,\vert\phi\rangle)=\arccos{(\sqrt{\vert\langle\phi\vert\psi\rangle\vert^2})}\).

For states \(\vert\psi\rangle\) and \(\vert\phi\rangle=\hat{U}_{\theta}\vert\psi\rangle\), the argument of the arccosine function becomes \(\sqrt{P(\theta)} =\sqrt{\vert\langle\psi\vert U_{
\theta}\vert\psi\rangle\vert^2}\). We note that, by choosing the appropriate measurement basis, quantum-state distinguishability reduces to a binomial distribution, where the 'yes' probability equals the probability that the final state projects onto the initial state. To compute the rate of change of quantum-state distinguishability with respect to \(\theta\), let us denote the probabilities as \(P(\theta)=P_1(\theta)\) and \(1-P(\theta)=P_2(\theta)\):
\begin{equation}\label{dlambda/dtheta}
\begin{split}
\frac{d\Lambda}{d\theta}&=-\frac{1}{2}\sqrt{\frac{1}{P(\theta)(1-P(\theta))}\left(\frac{d P(\theta)}{d\theta}\right)^2}\\
&=-\frac{1}{2}\sqrt{\sum_{m=1}^2\frac{1}{P_m(\theta)}\left(\frac{dP_m(\theta)}{d\theta}\right)^2}.
\end{split}
\end{equation}
The quantity is also the negative of the square root of the Fisher information for deciding \(\theta\) (See SM2). We can also define other metrics based on the statistical distance. One useful function we will use for discussion is \(\sin(\Lambda)\), which has been used in the classical version for object tracking \cite{Comaniciu}. Explicitly, it is given by:
\begin{equation}\label{sin(Lambd)}
    \sin(\Lambda) = \sqrt{1-\vert\langle\psi\vert\phi\rangle\vert^2}.
\end{equation}
In the domain, \(\Lambda\in[0,\pi/2]\), the new measure is monotonically increasing with \(\Lambda\). The term \(2\sin(\Lambda)^2\) is the pure state special case of the Bures metric\cite{Masahito, Bures}. Similarly, the measure \(\cos(\Lambda)=\vert\langle\psi\vert\phi\rangle\vert\) which is monotonically decreasing with \(\Lambda\) is the square root of the fidelity of a pure state\cite{Nielsen}. Both measures can be viewed as linearizations of quantum distinguishability.

With the measures properly constructed, we can now identify the most and least sensitive states with respect to a given unitary. Let us parametrize the unitary operation by \(\hat{U}=e^{-i\theta\hat{G}}\) where \(\hat{G}\) is the generator of the unitary operation. The phase \(\theta\) is the parameter of interest. We consider the derivative of \(\sin(\Lambda)\) in the asymptotic limit \(\theta\rightarrow 0\) so that it features the state sensitivity (See SM3):
\begin{equation}
    \frac{d}{d\theta}\sin(\Lambda) = \sqrt{\langle\psi\vert\hat{G}^2\vert\psi\rangle-\langle\psi\vert\hat{G}\vert\psi\rangle^2}.
\end{equation}
The most sensitive pure state for a single-parameter-dependent unitary operation, \(\vert\psi_{max}\rangle\) must satisfy:
\begin{equation}\label{sensor condition}
   \begin{split}
    \langle\psi_{max}\vert\hat{G}^2\vert\psi_{max}\rangle&=\max_{\vert\psi\rangle\in\mathcal{H}}\langle\psi\vert\hat{G}^2\vert\psi\rangle,\\
       \langle\psi_{max}\vert\hat{G}\vert\psi_{max}\rangle&=0.
   \end{split}
\end{equation}
Where the maximization is done over the whole Hilbert space \(\mathcal{H}\). The least sensitive pure state \(\vert\psi_{min}\rangle\) must satisfy:
\begin{equation}
    \langle\psi_{min}\vert\hat{G}^2\vert\psi_{min}\rangle=\langle\psi_{min}\vert\hat{G}\vert\psi_{min}\rangle^2.
\end{equation}
For polarization rotation or phase delay, it is most convenient to describe such a process in the angular momentum basis. Then, the generator of such process is \(\hat{G} =\boldsymbol{u}\cdot\boldsymbol{\hat{J}}\), where \(\boldsymbol{u}\) is the rotation axis and the \(\boldsymbol{\hat{J}}\) is the vector angular momentum operator. Suppose the quantum state contains N photons; the most sensitive state lies in the \(J=N/2\) subspace. For \(\boldsymbol{u}\) known, the sensitivity is \(J^2=N^2/4\). For unknown \(\boldsymbol{u}\), the maximal sensitivity is \(J(J+1)/3=(N^2+2N)/12\). Both exhibit the Heisenberg scaling of \(1/N\) (See SM4).

Now, we will consider the same problem from the perspective of error-detection criteria. Let \(\{\vert \bar{i}\rangle\}\) represent the logical states for the code space, \(\mathcal{C}\). Let \{\(\hat{E}_a\)\} be a basis for the errors. If we require \{\(\hat{E}_a\)\} be detectable, then the resultant state \(\hat{E}_a\vert\bar{j}\rangle\) has to be projected outside of the code space. In other words, the error cannot shuffle code words \(\vert \bar{i}\rangle\). We obtain the error detection condition:
\begin{equation}\label{detection}
    \langle\bar{i}\vert \hat{E}_a\vert\bar{j}\rangle = \delta_{ij}C_{a}.
\end{equation}
If we require that the error be both detectable and correctable, we arrive at the Knill-Laflamme condition \cite{Knill}:
\begin{equation}\label{correct}
    \langle\bar{i}\vert \hat{E}_a^\dagger \hat{E}_b\vert\bar{j}\rangle = \delta_{ij}C_{ab}.
\end{equation}
The physical meaning of Eq.(\ref{correct}) is that if a non-trivial error happened to the code word \(\vert \bar{j}\rangle\), the resultant state has to be projected outside of the code space and the error space formed by the resultant state from errors happening to all other code words \(\vert \bar{i}\rangle\). Such a requirement guarantees the ability to identify the nature of the error. Furthermore, it guarantees the existence of a recovery algorithm. Let the corresponding recovery operators be \(\{\hat{R}_r\}\). The efficiency of the error-correction code is characterized by the maximum distance between the recovered state and the original state after recovery. Such a quantity is called the error of the state\cite{Knill}:
\begin{equation}\label{errorrecovery}
\begin{split}
    \mathcal{E} &= \max_{\vert \psi\rangle\in C}\sum_{r,a}\vert (\hat{R}_r\hat{E}_a-\langle\psi\vert \hat{R}_r\hat{E}_a\vert\psi\rangle)\vert\psi\rangle\vert^2 \\&= \max_{\vert \psi\rangle\in C}\sum_{a,r}\langle \hat{E}_a^{\dagger}\hat{R}_{r}^{\dagger}\hat{R}_r\hat{E}_a\rangle-\langle \hat{R}_r\hat{E}_a\rangle\langle \hat{E}_a^{\dagger}\hat{R}_r^{\dagger}\rangle.
\end{split}
\end{equation}
If the recovery process is perfect and the code space is properly constructed, then the \(\mathcal{E}\) should be zero.

The most basic class of code is featured by the decoherence-free subspace, which is a code space constructed so that no active recovery operation is needed \cite{Lidar}. The errors to be considered are unitary operations. The code space consists of a degenerate common eigenstate of the error generator. The effect of the errors is to add the same phase to the code words. Thus, the net effect is just an undetectable global phase. In this case, when computing the Error, we have the recovery operation set to the identity, and the error becomes 
    \(\mathcal{E} =\max_{\vert\psi\rangle\in C}[ 1-\vert \langle \psi\vert\phi\rangle\vert^2]\),
where \(\vert\phi\rangle=\hat{U}\vert\psi\rangle\).
We note that this expression is identical to the square of state distinguishability in Eq. (\ref{sin(Lambd)}) defined by Wootters, Braunstein, and Caves \cite{Wootters, Cave}. The Error of code words in a decoherence-free subspace would be minimized to zero. If we design the worst possible code space, we aim to make the error as large as possible. We seek a state that maximizes the error. If we assume \(\theta\) is a very small parameter, then we can approximate the error as:
\begin{equation}\label{Error3}
    \mathcal{E} = \theta^2\max_{\vert\psi\rangle} [ (\langle\psi\vert \hat{G}^2\vert\psi\rangle-\langle\psi\vert \hat{G}\vert \psi\rangle^2)].
\end{equation}
The above is a reasonable approximation for small unitary operations, since it captures the key insight of decoherence-free subspaces: the best codewords are the eigenstates of the generator, yielding an error of zero. If one reverses the strategy for constructing decoherence subspaces and instead maximizes the error, then we arrive at the most sensitive sensor state, given by the condition Eq.(\ref{sensor condition}). 

Now, let us consider a more general use of the Knill-Laflamme condition for error detection and correction. We note that both detection and correction conditions require the code words to be projected out of the code space when a non-trivial error occurs. The effect of an error is to transform the state to an orthogonal state outside the code space. Then, by measuring whether the states are still in the code space, one can detect and possibly correct the error. In the following, we will use the error-detection conditions to see how parameter extraction can occur within the error-correction framework. Let us illustrate this point with a simple example. Let \(\{\vert \bar{i}\rangle\}\) be the logical qubit as before. Suppose the error is a single-qubit unitary, \(\exp\{-i\theta \hat{X}_k\}\), applied to the kth qubit. We can write \(\exp\{-i\theta \hat{X}_k\}=\cos(\theta)\hat{I}_k+i\sin(\theta)\hat{X}_k\). Thus, the discrete error set contains \(\{\hat{\mathcal{I}},\ \hat{X}_k\}\). Let \(\vert\psi\rangle\) be a state in a code space that satisfies the error detection condition \(\langle\bar{i}\vert \hat{X}_k\vert\bar{j}\rangle=\delta_{ij} C_k\), we find probability for the effect of the unitary error being trivial is \(P_k=\vert\langle \hat{U}_k\rangle\vert^2 = \cos^2(\theta)+\sin^2(\theta)C_k^2\). For \(C_k=1\), we have \(P=1\), so all errors are trivial. The code space is the decoherence-free subspace, where any state is completely insensitive to the error. For \(C_k=0\), there is a chance \(\cos^2(\theta)\) that the error has no effect, while a chance of \(\sin^2(\theta)\) that the state is sent out of the code space and is orthogonal to the initial state. The parameter \(\theta\) can be extracted by projecting the state after the error back to the initial state. That is to say the condition \(\langle\bar{i}\vert \hat{X}_k\vert\bar{i}\rangle=0\) allows the best extraction of \(\theta\). The same principle applies to an arbitrary unitary on a single qubit k. Then, we have \(\exp\{-i\theta\boldsymbol{u}\cdot\boldsymbol{\hat{\sigma}}_k\} = \cos(\theta)\hat{I}_k+i\sin(\theta)(\boldsymbol{u}\cdot\boldsymbol{\hat{\sigma}}_k)\). Any state in the code space that can detect or correct an arbitrary unitary single-qubit error, as long as it is not in the decoherence-free subspace, can work as a sensor for detecting the parameter \(\theta\). For the optimal detection, we additionally requires \(\langle\bar{j}\vert\boldsymbol{u}\cdot\boldsymbol{\hat{\sigma}_k}\vert\bar{j}\rangle=0\), such that the error \(\boldsymbol{u\cdot\hat{\sigma}_k}\) is non-trivial. The ability to extract an unknown parameter is associated with minimizing the probability that the error is trivial. Further, states in the error-detecting code space guarantee the extraction of the corresponding parameter.

To discuss quantum enhancement, we consider single-qubit unitary errors happening simultaneously to all qubits identically. Such processes are polarization rotation or phase delay. We are considering the process given by
\(
    \hat{U}_{\theta}= (\cos(\theta)\hat{\mathcal{I}}+i\sin(\theta)\boldsymbol{u}\cdot\boldsymbol{\hat{\sigma}})^{\bigotimes N}\).
Assuming the parameter \(\theta\) is small, so that when performing an expansion, we keep only \(\theta^2\) terms. We obtain:
\begin{equation}\label{Utheta}
\begin{split}
    & \hat{U}_\theta \approx \hat{\mathcal{I}}^{\otimes N}+i\theta\left(\sum_{k}\hat{\mathcal{I}}^{\otimes N-1}\otimes(\boldsymbol{u}\cdot\boldsymbol{\hat{\sigma}_k})\right)\\
    &-\theta^2\left(\frac{N}{2}\hat{\mathcal{I}}^{\otimes N}+\sum_{k,\ l>k}\hat{\mathcal{I}}^{\otimes N-2}\otimes(\boldsymbol{u}\cdot\boldsymbol{\hat{\sigma}_k})\otimes(\boldsymbol{u}\cdot\boldsymbol{\hat{\sigma}_l})\right).
\end{split}
\end{equation}
As long as \(\theta\) is small, we can approximate the error set as \(\{\hat{\mathcal{I}},\ \left(\boldsymbol{u}\cdot\boldsymbol{\hat{\sigma}_k}\right),\ \left(\boldsymbol{u}\cdot\boldsymbol{\hat{\sigma}_k}\right)\otimes\left(\boldsymbol{u}\cdot\boldsymbol{\hat{\sigma}_l}\right)\}\). Here, k and l denote the kth and the lth qubit. In other words, we approximated identical errors across all N qubits as single- and two-qubit errors affecting only part of the system. This ability for us to make the approximation is exactly where quantum enhancement comes from.

Let \(\vert\psi\rangle\) be an arbitrary state; the probability that the unitary error is trivial is
    \(P = 1-\theta^2\left[N+\sum_{k,\ l>k}2\langle\boldsymbol{u}\cdot\boldsymbol{\hat{\sigma}_k}\otimes\boldsymbol{u}\cdot\boldsymbol{\hat{\sigma}_l}\rangle-\left(\sum_{k}\langle\boldsymbol{u}\cdot\boldsymbol{\hat{\sigma}_k}\rangle\right)^2\right]\).
The parameter \(\theta\) can be found by inverting the above expression. Using the uncertainty propagation formula, the uncertainty in P is propagated into the uncertainty of \(\theta\) with a scaling factor given by \(
    \sqrt{N+\sum_{k,\ l>k}2\langle\boldsymbol{u}\cdot\boldsymbol{\hat{\sigma}_k}\otimes\boldsymbol{u}\cdot\boldsymbol{\hat{\sigma}_l}\rangle-\left(\sum_{k}\langle\boldsymbol{u}\cdot\boldsymbol{\hat{\sigma}_k}\rangle\right)^2}\).
For a state in the code space that can correct only a single qubit error associated with the discrete error set \(\{\hat{\mathcal{I}},\ \boldsymbol{u}\cdot\boldsymbol{\hat{\sigma}_k}\}\), the Knill-Laflamme condition requires:
\begin{equation}
\begin{split}
    \langle\bar{i}\vert\hat{\mathcal{I}}(\boldsymbol{u}\cdot\boldsymbol{\hat{\sigma_k}})\vert\bar{j}\rangle=\delta_{ij}C_{k},\\
    \langle\bar{i}\vert(\boldsymbol{u}\cdot\boldsymbol{\hat{\sigma}_k})(\boldsymbol{u}\cdot\boldsymbol{\hat{\sigma_l}})\vert\bar{j}\rangle=\delta_{ij}C_{kl}.\\
    .
\end{split}
\end{equation}
The scaling factor can be written in the coefficients \(C_k\) and \(C_{kl}\): \(
    \sqrt{N+\sum_{k,\ l\neq k}C_{kl}-\left(\sum_k C_k\right)^2} \). 
Often error correcting code are designed to additionally satisfy \(C_k=0\) and \(C_{kl} =\delta_{kl}\). In that case, we recover the shot-noise limit of \(\sqrt{N}\) scaling. These states provide no enhancement in estimating \(\theta\), but they have the capacity to estimate the magnitude of both the rotation axis and angle, and they serve as an error-correcting code for arbitrary single-qubit errors.

To maximize the scaling factor, we can relax the error-correcting requirement. We noticed the equivalence between the angular momentum operator and the Pauli operator when acting on states:
\begin{equation}
    \boldsymbol{u}\cdot\boldsymbol{\hat{J}}=\sum_m u_m\hat{J}_m \equiv \sum_k \sum_m u_m\hat{\sigma}_m^k=\sum_k \boldsymbol{u}\cdot\boldsymbol{\hat{\sigma}_k},
\end{equation}
\begin{equation}
    \left(\boldsymbol{u}\cdot\boldsymbol{\hat{J}}\right)^2 = N\hat{\mathcal{I}}+2\sum_{k,\ l>k} (\boldsymbol{u}\cdot\boldsymbol{\hat{\sigma}_k})\otimes(\boldsymbol{u}\cdot\boldsymbol{\hat{\sigma}_l}).
\end{equation}
The approximated unitary error up to \(\theta^2\) order is now:
\begin{equation}
    \hat{U}_{\theta} = \hat{\mathcal{I}}+i\theta\boldsymbol{u}\cdot\boldsymbol{\hat{J}}-\frac{\theta^2}{2}\left(\boldsymbol{u}\cdot\boldsymbol{\hat{J}}\right)^2.
\end{equation}
Correspondingly, we can write the qubit states in terms of angular-momentum states. Instead of trying to detect and correct single-qubit errors, we are now only interested in detecting or correcting single angular-momentum errors. For our particular unitary operation, the discrete error set is \(\{\hat{\mathcal{I}},\ \boldsymbol{u}\cdot\boldsymbol{\hat{J}}\}\). The physical states are now the angular momentum states, \(\{\vert J,\ m\rangle\}\), and the logical state is a linear combination of the physical states. This corresponds to a class of error-correcting codes designed to correct arbitrary changes in angular-momentum states. Such codes are referred to as absorption-emission (AE) codes. An example of a symmetric AE code space that corrects first-order error is given by \cite{Jain}. The non-trivial error probability for any state is \(
    P=1-\theta^2[\langle(\boldsymbol{u}\cdot\boldsymbol{\hat{J}})^2\rangle-\langle\boldsymbol{u}\cdot\boldsymbol{\hat{J}}\rangle^2]\).
If we do not know \(\boldsymbol{u}\), then the error set is \(\{\hat{\mathcal{I}},\ \hat{J}_x,\ \hat{J}_y,\ \hat{J}_z\}\equiv\{\hat{\mathcal{I}},\ \hat{J}_{+},\ \hat{J}_{-},\ \hat{J}_z\}\). If we assume \(\Delta m\geq2\) between the components of logical state \(\vert\bar{0}\rangle\) and \(\vert\bar{1}\rangle\). We automatically satisfied \(\langle\bar{i}\vert\hat{J}_{\pm}\vert\bar{j}\rangle=0\).
If we further assume the physical states are distributed equally around \(m=0\) in constructing the logical state, then \(\langle\bar{i}\vert\hat{J}_{z}\vert\bar{j}\rangle=0\). We can also show \(\langle\hat{J}_m\hat{J}_n\rangle=\delta_{mn}\langle\hat{J}_m^2\rangle\) (See SM4). The scaling factor is:
\(
\langle(\boldsymbol{u}\cdot\boldsymbol{\hat{J}})^2\rangle = \sum_{m}u_m^2\langle\hat{J}_m^2\rangle
\).
For the \(\Delta m\geq3\) case, one can further show \(\langle\hat{J}_x^2\rangle=\langle\hat{J}_y^2\rangle\). Since the angular momentum state is constrained by \(\langle\hat{J}^2\rangle=J(J+1)\) for the subspace labelled by J, constructing a second-order anti-coherent state is simple; we only need to check \(\langle\hat{J}_z^2\rangle=J(J+1)/3\). Then, we recover the Heisenberg limit for sensing polarization rotation, \(\langle(\boldsymbol{u}\cdot\boldsymbol{\hat{J}})^2\rangle =J(J+1)/3=(N/2)(N/2+1)/3\).

In conclusion, we derived the sensitivity of a pure state to a unitary change using the statistical distance defined by Wootters. We promoted this classical quantity to quantum state distinguishability, \(\Lambda\), which tells the difference between two pure quantum states. We explicitly showed that the Error of state, featuring the distortion of code words by the unitary, is a linearized quantitative description of quantum state distinguishability. When the change of parameter \(\theta\) is small
    \(\frac{d\Lambda}{d\theta}\approx\frac{d\sin(\Lambda)}{d\theta}=\sqrt{\langle\hat{G}^2\rangle-\langle\hat{G}\rangle^2}\), where \(\hat{G}\) is the generator of the unitary \(\hat{U}_\theta=\exp(-i\theta\hat{G})\). 
Using this relationship, one can maximize sensitivity and find the state that yields the largest Fisher information. We explored the connection between error-detection/correction codes and sensors and provided a sufficiency condition for constructing the second-order anti-coherent state for optimal rotation sensing. Specifically, we showed that to construct an optimal state for sensing a rotation with an unknown axis, we can use a quantum state with the form:
\(
    \vert\psi\rangle=\alpha_0\vert J,0\rangle+\sum_{m\geq1}\alpha_m\left(\vert J,m\rangle+\vert J,-m\rangle\right)\),
and having the coefficients \(\alpha_m\) satisfy \(\alpha_m^*\alpha_{m\pm1}=0\) and \(\alpha_m^*\alpha_{m\pm2}=0\) and satisfy \(\langle\psi\vert\hat{J}_z^2\vert\psi\rangle=J(J+1)/3\).
Our research provides compelling evidence that by leveraging principles from quantum error correction, one can significantly simplify sensor design.


\begin{thebibliography}{}
\bibitem{Huver}
Huver, Sean D. and Wildfeuer, Christoph F. and Dowling, Jonathan P., Entangled Fock states for robust quantum optical metrology, imaging, and sensing, Phys. Rev. A.78.063828,
(2008)
https://link.aps.org/doi/10.1103/PhysRevA.78.063828

\bibitem{Degen}
Degen, C. L. and Reinhard, F. and Cappellaro, P., Quantum sensing,
Rev. Mod. Phys.89.035002, (2017)
https://link.aps.org/doi/10.1103/RevModPhys.89.035002

\bibitem{Kaubruegger}
Kaubruegger, Raphael and Shankar, Athreya and Vasilyev, Denis V. and Zoller, Peter, Optimal and Variational Multiparameter Quantum Metrology and Vector-Field Sensing,
PRX Quantum.4.020333, (2023)
https://link.aps.org/doi/10.1103/PRXQuantum.4.020333

\bibitem{Pezze}
Pezz\`e, Luca and Smerzi, Augusto and Oberthaler, Markus K. and Schmied, Roman and Treutlein, Philipp, Quantum metrology with nonclassical states of atomic ensembles, Rev. Mod. Phys.90.035005, (2018)
https://link.aps.org/doi/10.1103/RevModPhys.90.035005

\bibitem{Reilly}
Reilly, Jarrod T. and Wilson, John Drew and J\ "ager, Simon B. and Wilson, Christopher and Holland, Murray J., Optimal Generators for Quantum Sensing, Phys. Rev. Lett.131.150802, (2023)
https://link.aps.org/doi/10.1103/PhysRevLett.131.150802

\bibitem{Giovannetti}
V. Giovannetti, S. Lloyd, and L. Maccone, Science 306, 1330
(2004).

\bibitem{Toth}
G. Tóth, Phys. Rev. A 85, 022322 (2012).

\bibitem{Goldberg2}
A. Z. Goldberg and D. F. V. James, Quantum-limited Euler angle measurements using anticoherent states, Phys. Rev. A 98, 032113 (2018)

\bibitem{Gunnar}
Gunnar Björk et al 2015 Phys. Scr. 90 108008

\bibitem{Baguette}
D. Baguette, F. Damanet, O. Giraud, and J. Martin, Phys. Rev. A 92, 052333 (2015).

\bibitem{Goldberg}
A. Goldberg, A. B. Klimov, G. Leuchs, and L.
L. Sánchez-Soto, Journal of Physics: Photonics
10.1088/2515- 7647/abeb54 (2021)

\bibitem{Cave}
S. L. Braunstein and C. M. Caves, Phys. Rev. Lett. 72, 3439
(1994).

\bibitem{Terhal}
Terhal, Barbara M., Quantum error correction for quantum memories,
Rev. Mod. Phys.87.307, (2025)
https://link.aps.org/doi/10.1103/RevModPhys.87.307

\bibitem{Li}
Li, Yaodong and Fisher, Matthew P. A., Statistical mechanics of quantum error correcting codes, Phys. Rev. B.103.104306,
https://link.aps.org/doi/10.1103/PhysRevB.103.104306

\bibitem{Jain}
Jain, Shubham P. and Hudson, Eric R. and Campbell, Wesley C. and Albert, Victor V., Absorption-Emission Codes for Atomic and Molecular Quantum Information Platforms, Phys. Rev. Lett.133.260601,
https://link.aps.org/doi/10.1103/PhysRevLett.133.260601

\bibitem{Knill}
Knill, Emanuel and Laflamme, Raymond, Theory of quantum error-correcting codes, Phys. Rev. A.55.900, (1997)
https://link.aps.org/doi/10.1103/PhysRevA.55.900

\bibitem{Fischer}
Kwon, H., Fischer, U.R., Lee, SW. et al. Restoring Heisenberg scaling in time via autonomous quantum error correction. npj Quantum Inf 12, 103 (2026). https://doi.org/10.1038/s41534-026-01268-1

\bibitem{Zhou}
S. Zhou and L. Jiang, Asymptotic Theory of Quantum Channel Estimation, PRX Quantum 2, 010343 (2021).

\bibitem{Sahu}
H Sahu, Q Xu, S Zhou, Achieving the Heisenberg limit using fault-tolerant quantum error correction, arXiv preprint arXiv:2601.05457 (2026)

\bibitem{Lidar}
Daniel A. Lidar, Review of Decoherence Free Subspaces, Noiseless Subsystems, and Dynamical Decoupling, Adv. Chem. Phys. 154, 295-354 (2014)

\bibitem{Wootters}
W. K. Wootters. Statistical distance and hilbert space. Phys. Rev. D, 23:357–362, Jan 1981.

\bibitem{Bhattacharyya}
Bhattacharyya, A. "On a Measure of Divergence between Two Multinomial Populations." Sankhyā: The Indian Journal of Statistics (1933-1960), vol. 7, no. 4, 1946, pp. 401–06. JSTOR

\bibitem{Nielsen}
M. A. Nielsen and I. L. Chuang, Quantum Computation and Quantum Information (Cambridge University Press, Cambridge, 2010), p. 400.

\bibitem{Comaniciu}
D. Comaniciu, V. Ramesh and P. Meer, "Kernel-based object tracking," in IEEE Transactions on Pattern Analysis and Machine Intelligence, vol. 25, no. 5, pp. 564-577, May 2003, doi: 10.1109/TPAMI.2003.1195991. keywords: {Target tracking;Cameras;Filters;Face detection;Filtering;Layout;State-space methods;Nonlinear equations;Kernel;Performance evaluation}

\bibitem{Masahito}
Masahito Hayashi, Quantum information theory, second ed., Graduate Texts in Physics, Springer-Verlag, Berlin, 2017, Mathematical foundation. MR 3558531

\bibitem{Bures}
Donald Bures, An extension of Kakutani's theorem on infinite product measures to the tensor product of semifinite w-algebras, Trans. Amer. Math. Soc. 135 (1969), 199–212. MR 236719

\bibitem{Cowan}
R.D. Cowan: The theory of atomic structure and spectra (University o 
 California Press, Berkeley, CA, 1981)
\end{thebibliography}
\end{document}